\documentclass[twocolumn,showpacs,preprintnumbers,amsmath,amssymb,aps,superscriptaddress]{revtex4-2}
\usepackage{graphicx}
\usepackage{dcolumn}
\usepackage{bm}
\usepackage[unicode=true,colorlinks=true,citecolor=blue,urlcolor=blue]{hyperref}
\usepackage{epstopdf}
\usepackage[T1]{fontenc} 
\usepackage{lmodern}
\usepackage{braket}
\usepackage{ulem}
\usepackage{color}

\renewcommand{\d}{\mathrm d}
\renewcommand{\i}{\mathrm i}
\newcommand{\e}{\mathrm{e}}
\newcommand{\eps}{\varepsilon}

\begin{document}

\title{Hyperfine versus exchange interaction in the spin dynamics\\ of spatially indirect excitons in CsPbI$_{3}$ perovskite nanocrystals}



\author{Timur S. Shamirzaev}
\email{tim@isp.nsc.ru}
\affiliation{Rzhanov Institute of Semiconductor Physics, Siberian Branch of the Russian Academy of Sciences, 630090 Novosibirsk, Russia}

\author{Nataliia E. Kopteva}
\affiliation{Experimentelle Physik 2, Technische Universit\"at Dortmund, 44227 Dortmund, Germany}

\author{Dmitry S. Smirnov}
\affiliation{Ioffe Institute, Russian Academy of Sciences, 194021 St. Petersburg, Russia}
\affiliation{Spin Optics Laboratory, St. Petersburg State University, 198504 St. Petersburg, Russia}

\author{Dmitri R. Yakovlev}
\affiliation{Experimentelle Physik 2, Technische Universit\"at Dortmund, 44227 Dortmund, Germany}
\affiliation{Ioffe Institute, Russian Academy of Sciences, 194021 St. Petersburg, Russia}

\author{Elena V. Kolobkova}
\affiliation{ITMO University, 199034 St. Petersburg, Russia}
\affiliation{St. Petersburg State Institute of Technology, 190013 St. Petersburg, Russia}

\author{Maria S. Kuznetsova}
\affiliation{Spin Optics Laboratory, St. Petersburg State University, 198504 St. Petersburg, Russia}

\author{Eugeniyus L. Ivchenko}
\affiliation{Ioffe Institute, Russian Academy of Sciences, 194021 St. Petersburg, Russia}

\author{Yan E. Maidebura}
\affiliation{Rzhanov Institute of Semiconductor Physics, Siberian Branch of the Russian Academy of Sciences, 630090 Novosibirsk, Russia}

\author{Evgeny A. Zhukov}
\affiliation{Experimentelle Physik 2, Technische Universit\"at Dortmund, 44227 Dortmund, Germany}

\author{Manfred Bayer}
\affiliation{Experimentelle Physik 2, Technische Universit\"at Dortmund, 44227 Dortmund, Germany}
\affiliation{Research Center FEMS, Technische Universit\"at Dortmund, 44227 Dortmund, Germany}

\begin{abstract}
We study the dynamics of recombination, optical orientation, and optical alignment of excitons in ensembles of CsPbI$_{3}$ nanocrystals (NCs), synthesized in a glass matrix. In large NCs with size exceeding 16~nm, the low-energy photoluminescence is contributed by the emission of indirect in real space excitons formed by spatially separated electrons and holes, which are localized at the NC/glass interface. The recombination dynamics of an ensemble of such excitons extends from tens of nanoseconds to microseconds and exhibits a power-law dependence. Their optical alignment and optical orientation reveal a peculiar spin dynamics caused by excitons influenced by the exchange interaction, varying by orders of magnitude. We develop a theory of the polarized photoluminescence of triplet excitons, taking into account the interplay between the electron-hole exchange interaction, their Zeeman effect, and their hyperfine interaction with the nuclei. This model reveals that for the excitons with the smallest exchange splitting we reach the regime, where the exciton fine structure becomes dominated by the hyperfine interaction with the random nuclear spin fluctuations in the NCs.
\end{abstract}
\maketitle
\section{Introduction}
\label{sec:intro}
Lead-halide perovskite nanocrystals (NCs) have emerged as a highly promising platform for modern optoelectronics~\cite{Vinattieri2021_book,Vardeny2022_book} and quantum photonics, owing to their outstanding optical properties~\cite{Kovalenko2017}, including high photoluminescence quantum yield~\cite{Sutter-Fella}, strong light--matter interaction, and efficient spin-dependent optical response~\cite{Vardeny2022_book,wang2019,ning2020,kim2021}. In particular, all-inorganic CsPbX$_3$ (X = Cl, Br, I) NCs demonstrate attractive features such as a tunable band structure, pronounced quantum confinement, and bright exciton emission~\cite{Protesescu2015}, making them appealing systems for investigating fundamental exciton physics.

The excitonic and spin-related phenomena in perovskite NCs have been intensively studied both experimentally and theoretically during the last years. Considerable progress has been achieved in understanding the exciton fine structure~\cite{Becker2018,Yin119}, exchange interaction~\cite{Fu2017}, exciton coherence~\cite{Gao2024,Han2022}, spin dynamics~\cite{nestoklon2023,Grigoryev2021,meliakov2025,meliakov2024,Kirstein2023}, spin noise~\cite{Kozlov2025}, and role of acoustic and optical phonons~\cite{Harkort2025,bataev2024,Trifonov2025,Zhu2025a,Lv2021} in these materials. In particular, polarization-resolved spectroscopy~\cite{He2025} and theoretical modeling have established the level structure of the bright and dark exciton states and clarified the role of crystal symmetry, confinement, and electron-hole exchange interaction in determining the optical selection rules and recombination dynamics~\cite{Sercel2019}. These aspects were addressed in numerous studies on ensembles of NCs and single NCs.

\begin{figure*}[hbt]
\centering
\includegraphics* [width=17cm]{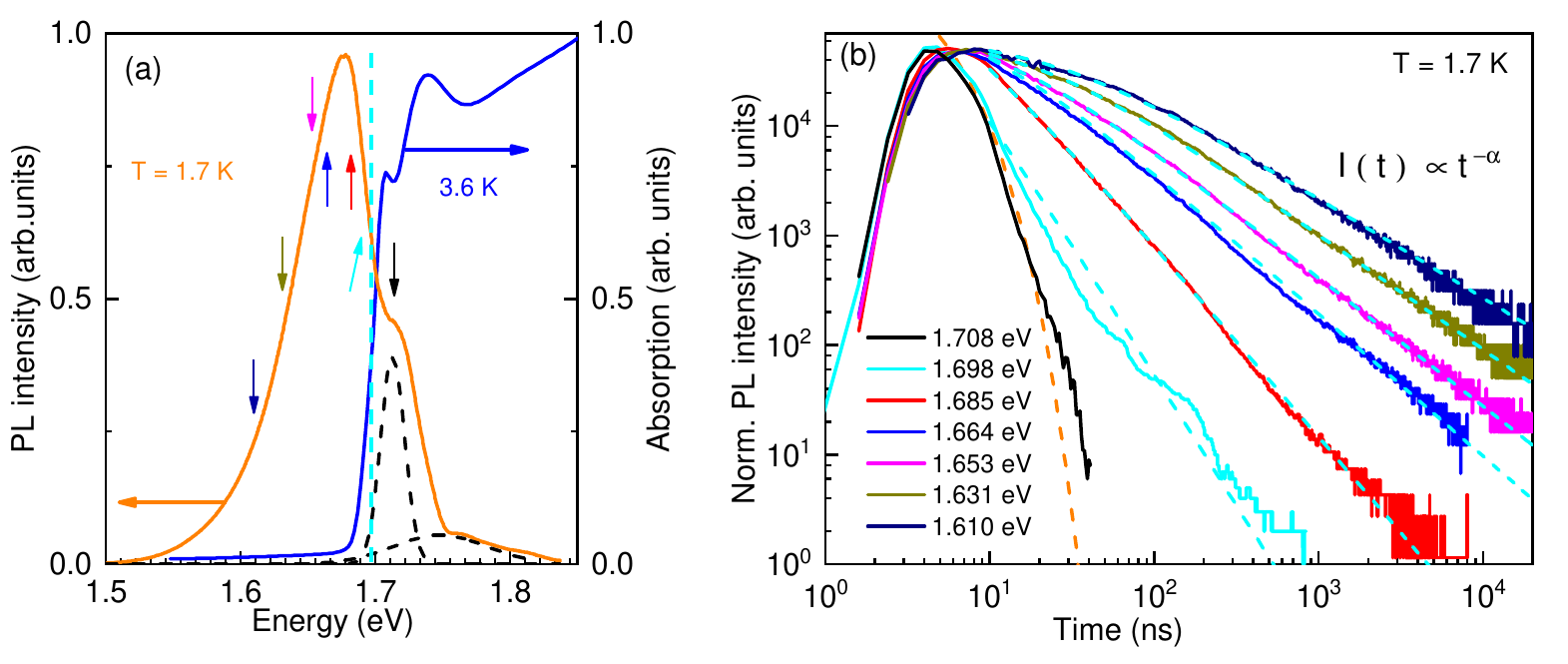}
\caption{ (a)  Optical absorption (blue line) and cw photoluminescence (orange line) spectra of CsPbI$_3$ NCs in a glass matrix. The vertical, cyan-dashed line marks the absorption edge at 1.698~eV. The black dashed lines highlight the two PL bands above the absorption edge at 1.708~eV and 1.739~eV. The vertical arrows denote the energies where the PL dynamics are measured with their colors corresponding to the lines in (b). (b) PL dynamics measured with nanosecond resolution at various energies in the range from 1.610 up to 1.708~eV. The orange dashed line shows an exponential fit which corresponds to the time resolution of the detection system. In order to accurately measure this dynamics, we used a streak camera that reveals the decay time of 730~ps (see Fig.~\ref{SIF0} in SI). The cyan dashed lines are fits using Eq.~\eqref{eq1}, implying a power aw decay with the parameters given in Table~\ref{tab:parameters}.
}
\label{Fig1}
\end{figure*}

At the same time, single-NC spectroscopy has reached a level, where the exciton spin structure and polarization properties can be studied with high precision~\cite{Hou2021,Tamarat2020,Tamarat2019,Fu2017}. However, most studies have focused on direct excitons, formed by strongly overlapping electron and hole wavefunctions confined within the NC volume. By contrast, spatially indirect excitons~\cite{Belykh2022} in perovskite NCs have received much less attention and remained largely unexplored. In particular, the role of spatial carrier separation for the exciton fine structure and spin dynamics has not yet been systematically investigated. For such excitons, the reduced overlap of the electron and hole wavefunctions is expected to strongly suppress the exchange interaction, thereby opening a regime where weaker spin interactions may become important~\cite{Shi2019,Liu2019,Chirvony2020,Becker2020,PhysRevB.108.195432}.

One of these interactions is the hyperfine interaction between charge carriers and nuclear spins, which is known to affect the spin dynamics of resident charge carriers (but not excitons) in both bulk lead-halide perovskites and perovskite NCs~\cite{Kirstein2023,hf2,Kotur2026_FAPI,Kotur2026_MAPI,kudlacik2024optical,Kirstein2023DNSS}. Previous studies demonstrated that fluctuating nuclear fields can substantially affect the spin polarization and spin coherence~\cite{hf3}. Nevertheless, the interplay between the hyperfine interaction and the electron-hole exchange interaction in perovskite NCs has remained poorly understood, especially in the regime, where the exchange splitting becomes strongly reduced.

In this work, we realize this regime by studying CsPbI$_3$ NCs embedded in a glass matrix. The synthesis conditions result in the formation of large NCs supporting spatially indirect excitons, each composed of an electron and a hole localized near the NC/glass interface. These excitons exhibit strongly suppressed exchange interaction and unusually long recombination dynamics, extending from tens of nanoseconds to microseconds. Their spin dynamics reveal a unique intermediate regime governed by the competition between exchange interaction, Zeeman splitting, and hyperfine interaction with fluctuating nuclear spins. This regime has never been reported before, to the best of our knowledge. By combining polarization-resolved spectroscopy with theoretical modeling, we demonstrate that for the excitons with the smallest exchange splitting, the fine structure becomes dominated by the random nuclear spin fluctuations in the NC ensemble.

\section{Experimental results}
\subsection{Optical absorption and photoluminescence}

The synthesis of the CsPbI$_3$ NCs in glass relies on the rapid cooling of a glass melt enriched with the components required for the formation of lead-halide perovskite NCs. Such a process often results in spatially inhomogeneous cooling conditions and may lead to the coexistence of NC subensembles with different mean sizes. Examples of this behavior for CsPbI$_3$ NCs can be found in Refs.~\cite{meliakov2025,meliakov2024}. The optical spectra presented below indicate that this is also the case for the sample studied here.

Figure~\ref{Fig1}(a) shows the absorption and photoluminescence (PL) spectra of the sample. The absorption spectrum shown by the blue line exhibits two pronounced features at 1.708~eV and 1.739~eV above the absorption edge, which manifests itself at 1.698~eV. The orange line in Fig.~\ref{Fig1}(a) shows the PL spectrum measured under continuous-wave, nonresonant excitation at 1.94~eV. The spectrum consists of three bands. The most intense, asymmetric band is Stokes-shifted relative to the absorption edge and has a maximum at 1.677~eV with a full width at half-maximum (FWHM) of approximately 60~meV. Two additional PL bands appear at 1.708~eV and 1.739~eV, coinciding with the maxima observed in the absorption spectrum. These bands can be fitted with Gaussian functions that have FWHMs of 15~meV and 80~meV, respectively, as shown by the black dashed lines in Fig.~\ref{Fig1}(a).

The high-energy PL bands at 1.739 eV and 1.708 eV  are associated with direct excitons formed by an electron and a hole confined by the NC potential and characterized by a strong overlap of their wavefunctions (see SI for details). As a result, these excitons undergo fast radiative recombination with lifetimes shorter than 1~ns, and they exhibit strong exchange interaction. The broad low-energy PL band is attributed to large NCs with diameters exceeding 16~nm (exceeding the exciton diameter in bulk CsPbI$_3$~\cite{Protesescu,Hou}), for which the charge-carrier confinement effects become weak.

The distinct origin of the PL bands above and below the absorption edge is further evidenced by their different recombination dynamics shown in Fig.~\ref{Fig1}(b). The black curve corresponds to the PL decay measured at 1.708~eV, above the absorption edge. The decay is monoexponential, and measurements with picosecond temporal resolution [see Fig.~\ref{SIF0} in the supplementary information (SI)] yield a lifetime of 730~ps, which is typical for excitons in lead-halide perovskite NCs~\cite{Qinghui,Kulebyakina,Canneson}. In contrast, at detection energies below the absorption edge, the PL dynamics becomes significantly slower, reaching the microsecond range at 1.610~eV. Moreover, the dynamics is nonexponential and follows a power law, $I(t) \propto t^{-\alpha}$, as illustrated by the remaining curves in Fig.~\ref{Fig1}(b). The initial 2~ns of these dynamics measured with high temporal resolution is shown in Fig.~\ref{SIF0}. The power-law decay indicates a superposition of multiple monoexponential decay channels with different lifetimes for excitons emitting at the same photon energy~\cite{Shamirzaev84}.

\subsection{Optical orientation and alignment of excitons}

\begin{figure}[]
\centering
\includegraphics[width=7cm]{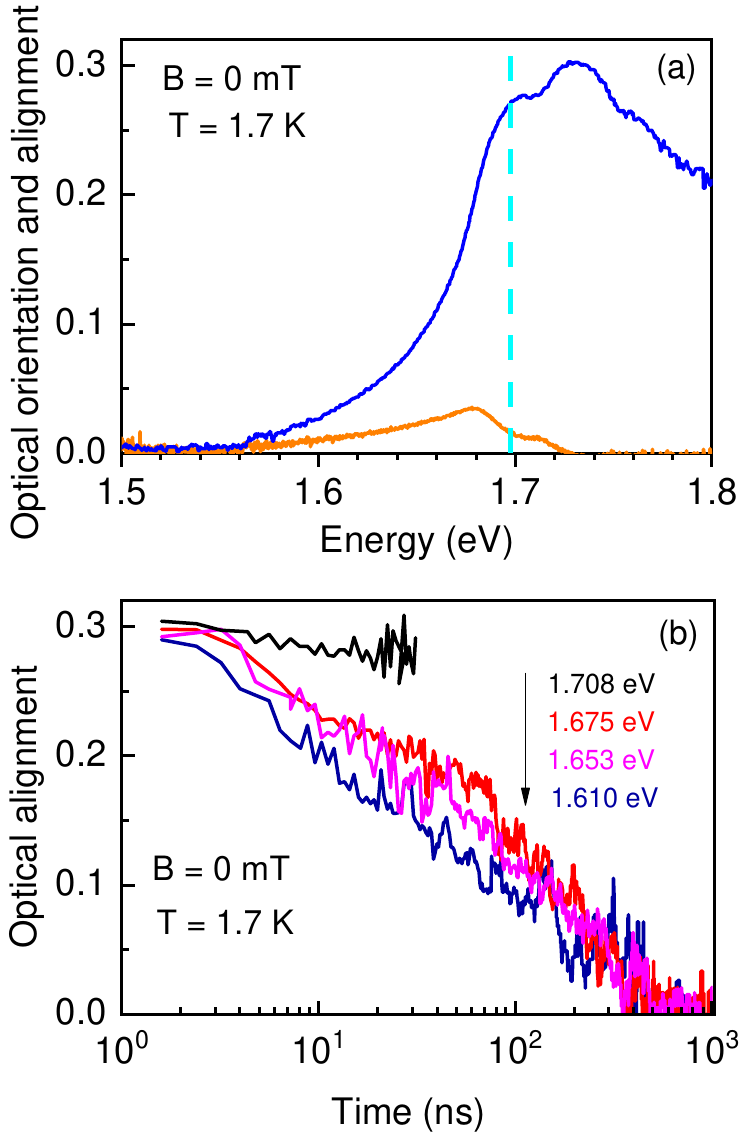}
\caption{\label{Fig2B0} Spectral dependence of (a) optical alignment (blue line) and optical orientation (orange line) for cw emission at zero magnetic field.  $T = 1.7$~K. The vertical cyan dashed line marks the absorption edge. (b) Dynamics of the optical alignment measured at energies of 1.708~eV (black), 1.675~eV (red), 1.653~eV (magenta), and 1.610~eV (blue).
}
\end{figure}

The fine structure and spin dynamics of excitons can be revealed by measuring the Stokes parameters of the PL polarization under polarized photoexcitation~\cite{OO,Dzhioev,Nestoklon,Astakhov,Dyakonov}. The orange line in Fig.~\ref{Fig2B0}(a) shows the spectrum of the degree of circular PL polarization measured under circularly polarized excitation, $P_\text{oo}$, corresponding to the optical orientation effect. $P_\text{oo}$ is zero in the high energy PL band at 1.739~eV, and remains small in the middle-energy band at 1.708~eV, where $P_\text{oo}$ reaches only about 1\%. In contrast, the strongest circular polarization, reaching 3.5\% is observed in the broad low-energy band below the absorption edge. This behavior indicates a specific exciton fine structure associated with the states contributing to the low-energy PL band.

The blue line in Fig.~\ref{Fig2B0}(a) shows the spectrum of the PL linear polarization degree ($P_{\rm l}$) under linearly polarized excitation. This parameter gives the optical alignment of excitons. We verified that the optical alignment is independent of the in-plane orientation of the excitation polarization, indicating that the NC ensemble is on average isotropic. In contrast to the optical orientation spectrum, the optical alignment is  strongest above the absorption edge, where it reaches 30\%. Below the absorption edge, $P_{\rm l}$ gradually decreases with decreasing energy. Notably, both the optical alignment and the optical orientation vanish at energies below 1.570~eV.

The dynamics of the optical alignment further highlight the fundamental difference between excitons emitting above and below the absorption edge, as shown in Fig.~\ref{Fig2B0}(b). At short time delays, the degree of linear polarization is approximately 30\% for all detection energies. At the above-edge energy, the polarization decreases only slightly with increasing delay, from 30\% to about 27\%, as illustrated by the black dynamics at 1.708~eV in Fig.~\ref{Fig2B0}(b). In contrast, below the absorption edge the optical alignment decays rapidly and vanishes at delays of about 500~ns. This pronounced difference indicates a qualitatively different nature of the excitons responsible for the PL bands below and above the absorption edge.

\subsection{Effect of magnetic field on exciton spin}

\begin{figure*}[hbt]
\centering\includegraphics*[width=0.8\linewidth]{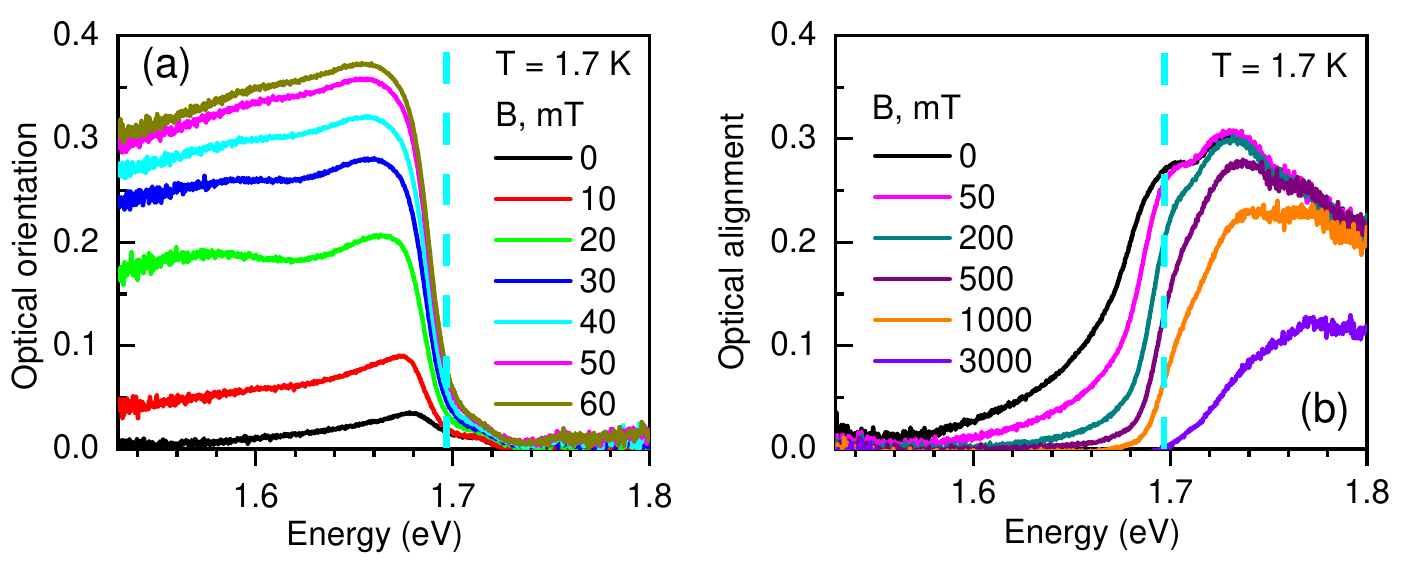}
\caption{\label{Fig3} Spectral dependences of optical orientation (a) and optical alignment (b), measured for cw emission in different longitudinal magnetic fields.  $T = 1.7$~K. The vertical dashed line marks the absorption edge.}
\end{figure*}

The details of the exciton fine structure can be investigated by applying an external magnetic field, which induces a Zeeman splitting of the exciton states~\cite{Dzhioev,ivchenko05a}. In particular, when the magnetic field is applied parallel to the optical axis (Faraday geometry), the optical orientation and alignment spectra change, as shown in Fig.~\ref{Fig3}. The optical orientation is rapidly recovered in weak magnetic fields of about 60~mT, increasing from a few percent up to approximately 35\% at spectral energies below the absorption edge, as shown in Fig.~\ref{Fig3}(a). In contrast, above the absorption edge, the optical orientation remains very small. The optical alignment dominates above the absorption edge and gradually decreases with increasing magnetic field [Fig.~\ref{Fig3}(b)]. However, the characteristic fields in this case reach several Tesla and are therefore much stronger than those required for the recovery of the optical orientation.

\begin{figure}[hbt]
\centering
\includegraphics*[width=8.5cm]{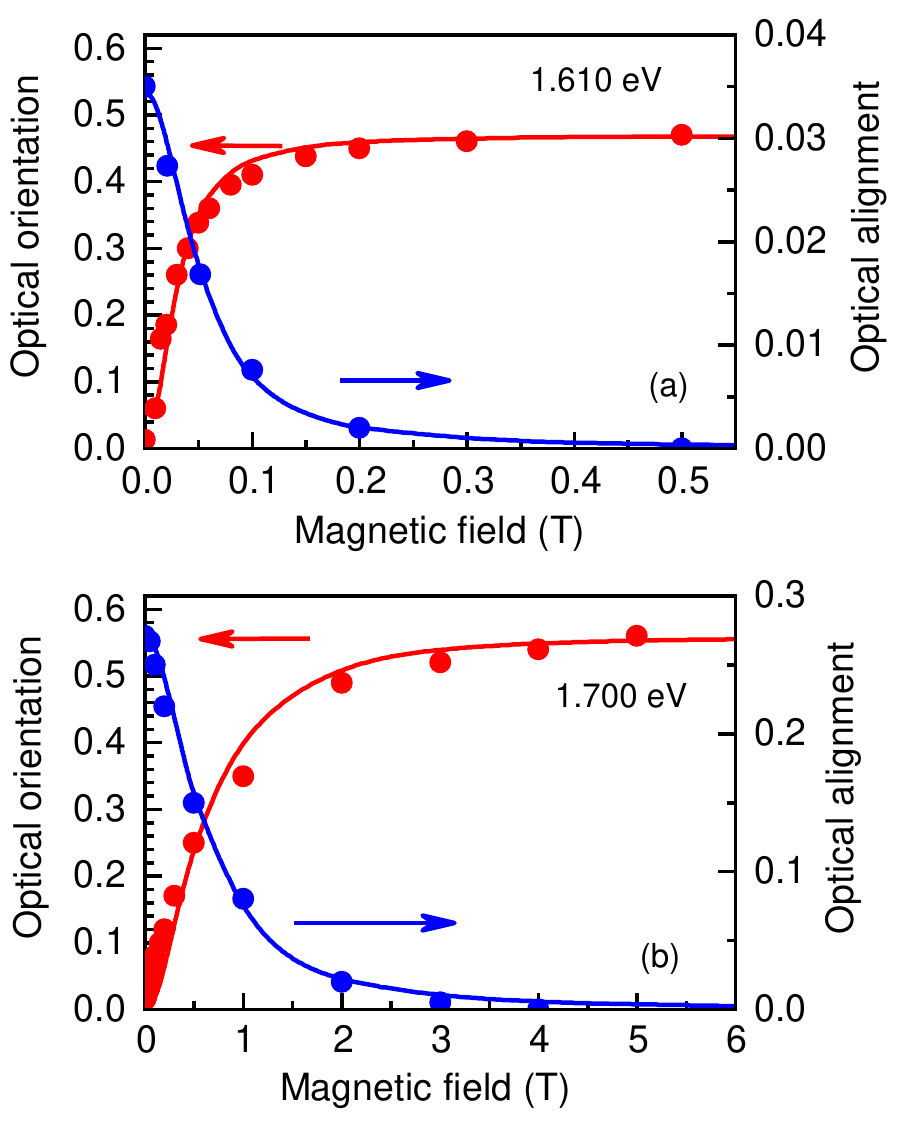}
\caption{\label{Fig4} Optical orientation and optical alignment for cw emission $vs$ longitudinal magnetic field measured at energies below (a) and above (b) the absorption edge. The lines are the results of model calculations with Eqs.~\eqref{eq:a1} and~\eqref{eq:a2} with the anisotropic exchange splitting $\delta$ and $P^{\text{sat}}_\text{oo}$, and $P_{\text{l}}^0$ values given in Table~\ref{tab:HWHM}.
}
\end{figure}

The magnetic field-induced changes are shown in more detail in Fig.~\ref{Fig4} for two detection energies: (a) below the absorption edge at 1.610~eV and (b) above the edge at 1.700~eV. One can see in Fig.~\ref{Fig4}(a) that below the absorption edge weak magnetic fields in the range of several tens of millitesla enhance the optical orientation up to 47\%, while simultaneously suppress the optical alignment from 3.5\% to zero. Such pronounced changes in the polarized luminescence induced by weak magnetic fields indicate small fine structure splittings and a strongly suppressed exchange interaction for the excitons emitting below the absorption edge.

In contrast, above the absorption edge, similar changes in the optical orientation and alignment require much stronger magnetic fields on the order of 1~T, as shown in Fig.~\ref{Fig4}(b). Such behavior is expected for direct excitons in NCs due to the interplay between the strong electron-hole exchange interaction and the Zeeman effect~\cite{Dzhioev,ivchenko05a}.

\begin{figure*}[hbt]
\centering
\includegraphics* [width=17cm]{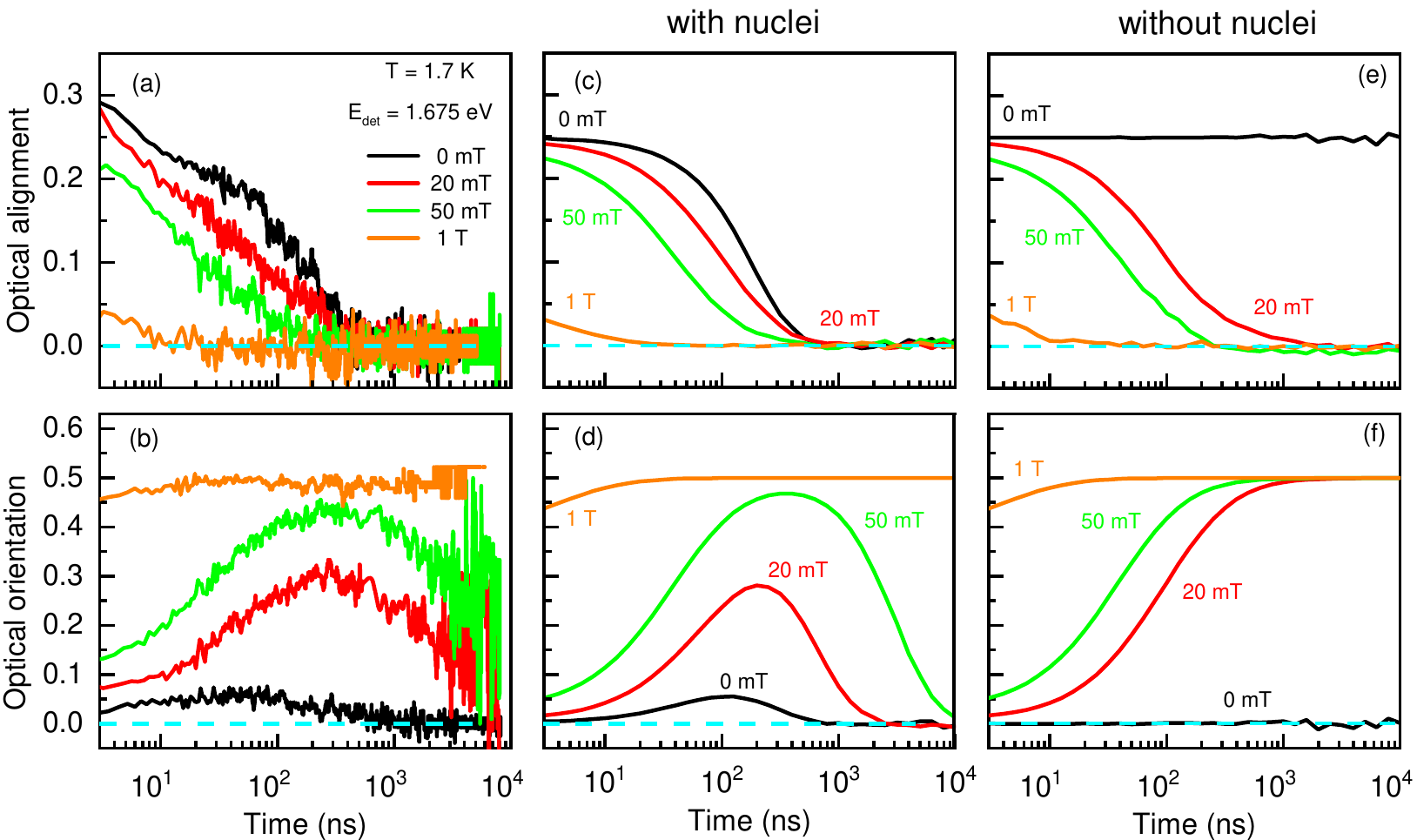}
\caption{\label{Fig5}  Dynamics of optical alignment (a) and  optical orientation (b) measured at the energy of 1.675~eV in different magnetic fields. Calculation of the polarization dynamics with account of the hyperfine interaction of the SIE with the nuclear spins (c,d) and without accounting for it (e,f). The model parameters are given in the main text.
     }
\end{figure*}

To obtain deeper insight into the spin dynamics of the excitons below the absorption edge, we study the dynamics of their optical orientation and alignment in different magnetic fields. As one can see in Figs.~\ref{Fig5}(a,b), increasing the magnetic field suppresses the optical alignment and enhances the optical orientation at all time delays. The optical alignment generally decreases with time and vanishes at long delays. In contrast, the optical orientation shows a nonmonotonous behavior: an initial rise is followed by a decay, with the growth occurring in antiphase with the suppression of the optical alignment. In strong magnetic fields approaching 1~T the optical orientation dynamics flattens and saturates at about 50\%.

The pronounced differences in the PL dynamics and the polarization properties between excitons below and above the absorption edge, as well as the peculiar magnetic field dependence of the dynamics of the optical orientation and alignment, call for a theoretical interpretation.

\section{Discussion and theory}
\subsection{Spatially indirect excitons}

The low-energy PL band with maximum at 1.677~eV is in the focus of our study. It is Stokes-shifted from the absorption edge at 1.698~eV, which is determined by direct excitons in large NCs with diameters exceeding 16~nm. The low-energy tail of this PL band extends over 100~meV and demonstrates a very long-lived, non-monoexponential recombination dynamics, see Fig.~\ref{Fig1}(b).  This allows us to conclude that this band is formed by the excitons in  large NCs, where at least one of the charge carriers is trapped by a localized state. Such localized states are known to appear at the NC/glass interfaces in perovskite NCs~\cite{Kulebyakina,Belykh2022,Belykh2026_ODMR_CsPbI3}. We suggest that the localization of an electron or a hole (or both) shifts the exciton energy below the absorption edge. The shape of this band reflects the energy distribution of the localized states in NCs rather than the distribution of their sizes.  Localization of charge carriers strongly reduces the overlap of the electron and hole wavefunctions in excitons, to which we coin the term of ``spatially indirect excitons'' (SIE). Their essential feature is the strongly reduced exchange interaction, which results in modified PL polarization and dynamics.

\subsection{Modeling of recombination dynamics}

For the spatially indirect excitons emitting below the edge, the PL decay is non-monoexponential, indicating the existence of many different species with different decay times $\tau$. This evidences a broad scatter in the overlap of the electron and hole wave functions in the spatially indirect excitons. Phenomenologically, this distribution can be described in the form~\cite{Shamirzaev84}
\begin{equation}
  \label{eq2}
  G(\tau) \propto \frac{1}{\tau^{\alpha+1}}
  \textnormal{exp}\left(-\frac{\tau_{0}}{\tau}\right),
\end{equation}
where $\tau_0$ is the typical exciton lifetime in the distribution, and $\alpha$ gives the PL decay $I(t) \propto t^{-\alpha}$ at long times. Generally, the PL decay is described by
\begin{equation}
  \label{eq1}
  I(t) = \int_0^{\infty} G(\tau) \exp \left( -\frac{t}{\tau} \right)\d\tau.
\end{equation}
We use this approach to fit the experimental dynamics in Fig.~\ref{Fig1}(b). The evaluated parameters of the distribution $\alpha$ and $\tau_0$ for the different detection energies are given in Tab.~\ref{tab:parameters}. The corresponding distribution functions~\eqref{eq2} are shown in Fig.~\ref{SIF1} in the SI.

\begin{table}[hbt]
  \small
  \caption{Parameters of the exciton lifetime distribution~\eqref{eq2},  obtained from fits of the PL decays in Fig.~\ref{Fig1}(b).}
  \label{tab:parameters}
  \begin{tabular*}{0.48\textwidth}{@{\extracolsep{\fill}}lll}
    \hline
    PL energy (eV)  & $\alpha$ &  $\tau_{0}$ (ns) \\
        \hline
    1.698   &  2.40   & $1\pm0.12$   \\
    1.682   &  1.75  & $2\pm0.15$   \\
    1.664   &  1.30  & $6\pm0.40$   \\
    1.653   &  1.17 & $9.5\pm0.40$   \\
    1.631   &  1.05 & $17\pm0.40$   \\
    1.610   &  0.90  & $23\pm0.50$  \\
    \hline
  \end{tabular*}
\end{table}

Note that even at a given, fixed detection energy, excitons with very different electron-hole overlaps coexist. From Tab.~\ref{tab:parameters} one can see that at the highest energy, the characteristic exciton lifetime is similar to the lifetime of the spatially direct excitons and the distribution function is quite sharp. For lower energies, the distribution shifts to  longer times and becomes broader: the deeper the localizing potential, the stronger the separation of electron and hole, as expected.

\subsection{Interplay of exchange and Zeeman interactions}

We now turn to the description of optical orientation and alignment of the SIE accounting for the external longitudinal magnetic field $\textbf{B}$. In the lead halide perovskite semiconductors, the extrema of the conduction and valence bands are two fold degenerate with respect to the electron and hole spins of 1/2~\cite{Kirstein_NC_2022}.  Accordingly, the exchange interaction splits the exciton states into a dark singlet and three bright triplet states~\cite{Nestoklon,Zhiliakov2026}.    Furthermore, the electron-hole long-range exchange interaction splits the bright excitons into three orthogonal linearly polarized states $X'$, $Y'$ and $Z'$ optically active for light polarized along the corresponding axes. These splittings are caused by various imperfections such as shape anisotropy, strain, randomly positioned defects, low symmetry crystal phases, etc. All of them can lead to the fine structure splitting via the electron-hole exchange interaction. Note that the states represent linear combinations of the states $X$, $Y$, and $Z$, which are polarized along the axes of the laboratory coordinate frame with the $z$ axis parallel to the optical axis.

Earlier theoretical investigations of the polarized luminescence of excitons in semiconductors with simple band structure were performed for bulk GaSe crystals~\cite{1977ZhETF2230I,1977JETP590G,Ivchenko_GaSe}. Later studies addressed CuCl nanocrystals~\cite{PhysRevB.47.16024,https://doi.org/10.1002/pssc.200879872,PhysRevB.111.085423}. The exciton fine structure in perovskite nanocrystals and its manifestations in the polarized photoluminescence in magnetic field was studied in Refs.~\cite{Nestoklon,Lounis,Tamarat18,Gao2024,Tamarat14,Yin11}. Recently, a theory accounting for the random orientation of the nanocrystals in the glass matrix and the competition the between Zeeman effect and the exchange interaction was developed~\cite{random_triplet}.

In general, the fine structure of the bright exciton triplet in perovskite NCs is determined by five parameters: the two splittings between the exciton states and the three Euler angles $\varphi, \theta, \psi$ specifying the directions of the linear polarization of the eigenstates. This is in sharp contrast with conventional nanostructures, where the doublet of the bright exciton states is characterized by the single splitting between the states and the single azimuthal angle of their polarization direction~\cite{ivchenko05a}.

The Hamiltonian of the bright excitons can be written in terms of the exciton angular momentum operator $\bm L$ ($L=1$) as
\begin{equation}
  \mathcal H_0=\bm L\hat{\bm\Delta}\bm L+g\mu_{\rm B}BL_z.
  \label{eq:Hb}
\end{equation}
Here $g$ is the bright exciton $g$-factor, and $\mu_{\mathrm B}$ is the Bohr magneton. in line with the experiments, we consider the Faraday geometry, and $\hat{\bm\Delta}$ is the tensor of the exchange splittings. In the basis of the exciton states $X$, $Y$, and $Z$, the tensor $\hat{\bm \Delta}$ is described by a Gaussian orthogonal ensemble of random matrices $3\times3$~\cite{mehta2004random} with the dispersion $\delta^2$ of the diagonal elements and $\delta^2/2$ of the off-diagonal elements. Thus, $\delta$ is, by definition, the typical scale of the exciton fine structure splitting.

In the realistic limit of long SIE lifetime, $\delta\tau/\hbar\gg 1$,  the PL linear polarization degree under linearly polarized excitation along the $x$ axes is given by
\begin{subequations}
  \label{eq:long}
  \begin{equation}
    \label{eq:l}
    P_\text{l}=\frac{\sum_{m}\left|\psi_{x}^{(m)}\right|^2\left(\left|\psi_{x}^{(m)}\right|^2-\left|\psi_{y}^{(m)}\right|^2\right)}{\sum_{m}\left|\psi_{x}^{(m)}\right|^2\left(\left|\psi_{x}^{(m)}\right|^2+\left|\psi_{y}^{(m)}\right|^2\right)},
  \end{equation}
where $\psi^{(m)}$ with $m=1,2,3$ are the three-component eigencolumns of the Hamiltonian~\eqref{eq:Hb} in the $X,Y,Z$ basis. The subscript denotes its components, accordingly. Similarly, the optical orientation degree under circularly polarized excitation is given by
  \begin{equation}
    \label{eq:oo}
    P_\text{oo}=\frac{\sum_{m}\left|\psi_{+}^{(m)}\right|^2\left(\left|\psi_{+}^{(m)}\right|^2-\left|\psi_{-}^{(m)}\right|^2\right)}{\sum_{m}\left|\psi_{+}^{(m)}\right|^2\left(\left|\psi_{+}^{(m)}\right|^2+\left|\psi_{-}^{(m)}\right|^2\right)},
  \end{equation}
\end{subequations}
where the eigencolumns are written in the basis of the states $\ket{L_z}$ with the angular momentum $L_z=0,\pm1$ along the $z$ axis. Explicitly, $\ket{\pm 1}=\left(\ket{X}\pm\i\ket{Y}\right/\sqrt{2}$ and $\ket{0}=\ket{Z}$.

The optical alignment and orientation calculated after Eqs.~\eqref{eq:long} are functions of the single dimensionless parameter $g\mu_{\rm B}B/\delta$. In the SI we numerically demonstrate that with a high accuracy they can be approximated by Lorentzians, similarly to the case of doublet excitons~\cite{Dzhioev,ivchenko05a}:
\begin{subequations}
  \label{eq:a}
  \begin{equation}
    \label{eq:a1}P_\text{l}(B) = P^{0}_\text{l}
    \frac{\delta^{2}}{\delta^{2}+(g\mu_{\rm B}B)^2} ,
  \end{equation}
  \begin{equation}
    \label{eq:a2}P_\text{oo}(B) = P^{\text{sat}}_\text{oo} \frac{(g\mu_{\text{B}}
      B)^2}{\delta^{2}+(g\mu_{\text{B}} B)^2} ,
  \end{equation}
\end{subequations}
where $P^{0}_\text{l}$ and $P_{\text{oo}}^{\text{sat}}$ represent the optical alignment at zero magnetic field and the optical orientation saturation level in strong magnetic fields.

\begin{figure*}[hbt]
\centering
\includegraphics[width=17cm]{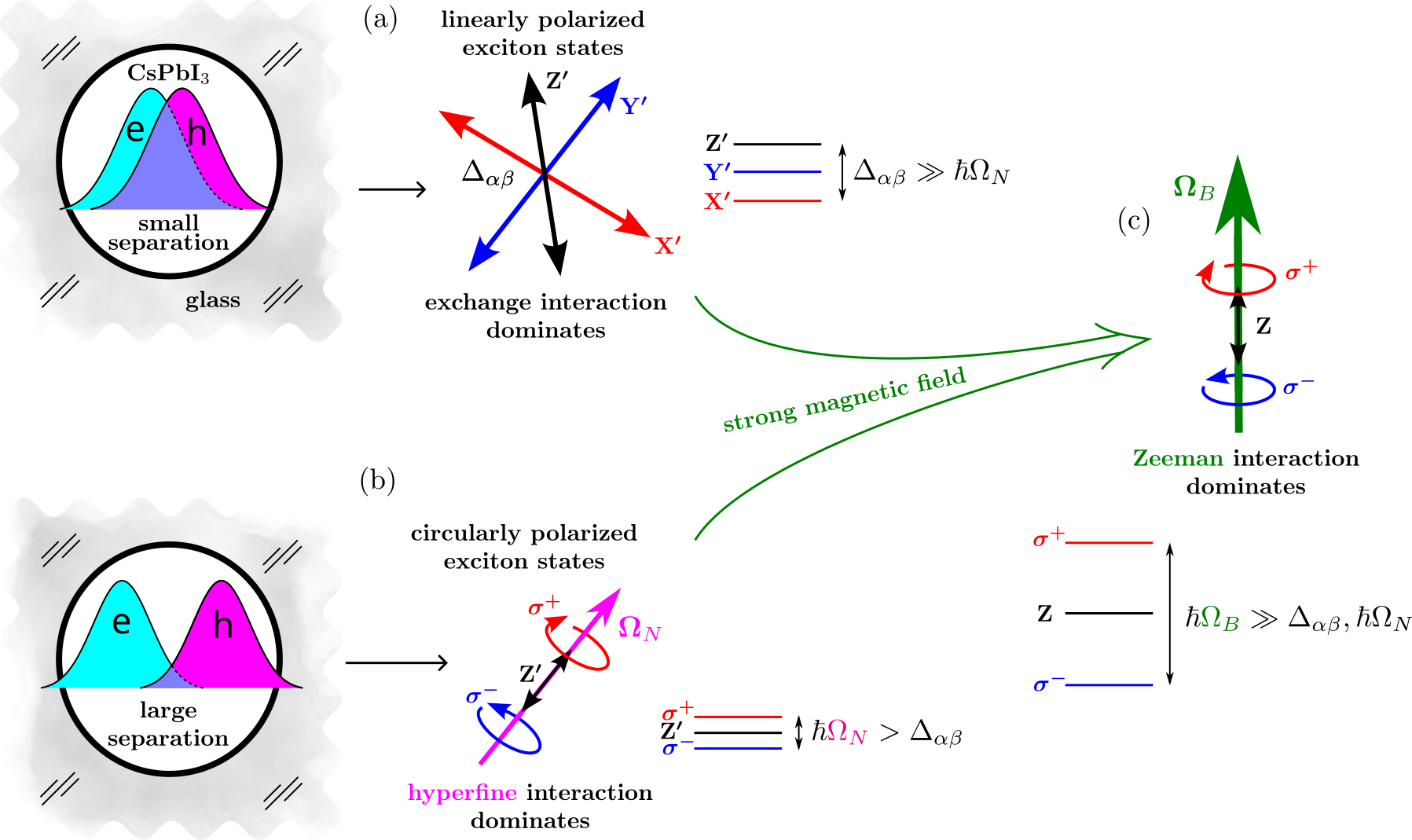}
\caption{Schematic illustration of CsPbI$_3$ NCs with spatially indirect excitons. The overlap of the electron and hole wave functions (purple area) varies strongly, which determines both the strength of the exchange interaction and the exciton recombination time. (a) Strong overlap of the wave functions is provided by a small separation. Here the exciton states are linearly polarized. (b) Weak overlap of the wave functions is provided by a larger separation. Here the exciton states are circularly polarized due to the hyperfine interaction. (c) Application of a magnetic field leads to the formation of circularly polarized SIE eigenstates with the projections of the angular momentum $0,\pm1$ onto the $z$ axis.
}
\label{Fig6}
\end{figure*}

The qualitative explanation of the magnetic field dependences is sketched in Fig.~\ref{Fig6}. Panel (a) shows a SIE in a perovskite NC and its linearly polarized eigenstates $X'$, $Y'$, $Z'$ in zero magnetic field. In the limit of long exciton lifetimes, the PL can be calculated separately for all exciton eigenstates, and the interference between these states can be neglected, as seen from Eqs.~\eqref{eq:long}. Due to the linear polarization of the eigenstates, the optical orientation of excitons vanishes, in agreement with Eq.~\eqref{eq:a2}. Application of a longitudinal magnetic field leads to the formation of the circularly polarized states $\ket{L_z}$ shown in Fig.~\ref{Fig6}(c). With increasing magnetic field the optical orientation reaches its maximum of $P^{\text{sat}}_\text{oo}$, while the optical alignment vanishes due to the circular polarization of the eigenstates, in agreement with Eqs.~\eqref{eq:a}.

We apply Eqs.~\eqref{eq:a} to fit the measured dependences, see the examples of such fits in Fig.~\ref{Fig4} and Fig.~\ref{SIF4} of the SI. Making use of the exciton $g$-factor $g=2.2$~\cite{meliakov2025,meliakov2024}, we obtain the values of $\delta$ corresponding to the measurements of optical orientation $\delta_{\rm oo}$ and alignment $\delta_{\rm l}$.  They are given in Tab.~\ref{tab:HWHM} for several detection energies.

\begin{table}[hbt]
\small
\caption{Parameters of the magnetic field dependences of optical orientation and alignment measured at different energies, fitted with Eqs.~\eqref{eq:a}.
 }
  \label{tab:HWHM}
  \begin{tabular*}{0.48\textwidth}{@{\extracolsep{\fill}}llllll}
    \hline
    PL energy (eV)                    & 1.576         & 1.610         & 1.660        & 1.675          & 1.700      \\    \hline
    $\delta_{\rm l}$   ($\mu$eV)      & 5.8$~\pm$0.2  & 6.3$~\pm$0.2  & 8.9$~\pm$0.2 & 9.3$~\pm$0.1   &  74$~\pm$1  \\
    $P_{\text{l}}^\text{0}~($\%$)$    & 1.5           & 3.5           & 10.6         & 18.0           &  27.0       \\
    $\delta_{\rm oo}$   ($\mu$eV)     & 3.1$~\pm$0.2  & 3.7$~\pm$0.2  & 3.9$~\pm$0.2 & 3.9$~\pm$0.1   &  72$~\pm$1  \\
    $P^{\text{sat}}_\text{oo}~($\%$)$ & 39            & 47            & 50           & 51             &  56          \\
           \hline
  \end{tabular*}
\end{table}

For the above-edge energy of 1.700~eV, $\delta_{\rm l}$ and $\delta_{\rm oo}$ are close to each other ($\sim73~\mu$eV) and agree well with the typical splitting of $120~\mu$eV, reported in Ref.~\cite{Nestoklon} for colloidal CsPbI$_3$ NCs at the PL detection energy of 1.720~eV. This confirms the interpretation of these states as direct excitons.

At lower energies, $\delta_{\rm l}$ and $\delta_{\rm oo}$ become much smaller, which evidences a weaker exchange interaction in the SIE due to the reduced overlap between the electron and hole wave functions.

The widths of the magnetic field dependence of optical orientation and alignment at lower energies differ approximately by a factor of 2. This is related to the fact that the dominant contribution to the optical alignment occurs at times earlier than that to the optical orientation. This can be seen in Fig.~\ref{Fig5}. Accordingly, the overlap between the electron and hole wave functions, as well as the exchange interaction, are stronger in the former case. This results in a broader magnetic field dependence of the optical alignment than of the optical orientation.

\subsection{Manifestation of hyperfine interaction}

The modeling of the optical alignment and orientation in conventional nanostructures neglects the distribution of the overlap between the electron and hole wave functions, which determines both the exciton radiative lifetime and the strength of the exchange interaction~\cite{BP,qdq1-2yk9}. However, for the SIE in CsPbI$_3$ NCs this distribution is broad and may play a significant role in the dynamics illustrated in Fig.~\ref{Fig5}. Moreover, with increase of the spatial separation between electron and hole, their exchange interaction can become weaker than the strength of the hyperfine interaction with the random nuclear spin fluctuations in NCs, which is typically of the order of 1~$\mu$eV~\cite{hf1,hf2,hf3}. In this case, one can reach a new regime, where the exciton fine structure is determined by the exciton hyperfine interaction with the nuclear spin fluctuations.

All these contributions can be taken into account following Ref.~\cite{random_triplet} by considering the following non-Hermitian SIE Hamiltonian:
\begin{equation}
  \mathcal H=\mathcal H_0+\hbar \bm L \bm\Omega_N-\i\hbar/(2\tau).
  \label{eq:H}
\end{equation}
Here $\bm\Omega_N$ is the exciton spin precession frequency in the random effective nuclear magnetic field (Overhauser field), which is normally distributed:~\cite{book_Glazov,PhysRevResearch.5.L032032}
\begin{equation}
  \label{eq:F}
  \mathcal F(\bm\Omega_N)=\left(\sqrt{\frac{2}{\pi}}T_2^*\right)^3\e^{-2\left(\Omega_NT_2^*\right)^2},
\end{equation}
with the exciton spin dephasing time $T_2^*$. The corresponding typical field of the nuclear spin fluctuations is $\Delta_B=\hbar/(\sqrt{2}T_2^*g\mu_{\rm B})$. We take into account the nuclear spin dynamics by the Markov switching between all the values in the distribution~\eqref{eq:F} with the correlation time $\tau_c$~\cite{Hopping_Glazov,PRC,tauc}, which is much longer than $T_2^*$. Note that the same Hamiltonian \eqref{eq:H} is also relevant for the spatially direct excitons, but the hyperfine interaction for them is much smaller than the exchange interaction, so the second term can be omitted. For the SIE this is not the case.

The optical alignment can be described similarly to Eq.~\eqref{eq:l} by solving the Schr\"odinger equation with the Hamiltonian~\eqref{eq:H} and the initial condition $\psi_\textit{i}(t)=\delta_{i,x}$ ($i=x,y,z$), which corresponds to polarization of the excitation along the $x$ axis.
Then one obtains for the intensities of the light emitted along the $z$ axis with polarization parallel to the $x$ or $y$ axes
\begin{equation}
  \label{eq:Ixy_tau}
  I_{x,y}(t)\propto\left|\psi_{x,y}(t)\right|^2/\tau.
\end{equation}
The degree of linear polarization is given by
\begin{equation}
  \label{eq:rhol}
  P_\text{l}(t)=\frac{\braket{I_x(t)}-\braket{I_y(t)}}{\braket{I_x(t)}+\braket{I_y(t)}},
\end{equation}
where the angular brackets denote averaging over $\hat{\bm\Delta}$, $\delta$, $\tau$ and $\bm\Omega_N$.  Note that since both the lifetime and the strength of the exchange interaction are determined by the overlap between the electron and hole wave functions, they are inversely proportional to each other: $\delta=\hbar\varkappa/\tau$, with a proportionality constant $\varkappa$.

In the same way, one can calculate the optical orientation by using the circular basis instead of the linear one. The corresponding initial condition for the wave function is $\psi_{m}(0)=\delta_{m,1}$ with $m$ being $0,\pm 1$ and the components $x,y$ in Eq.~\eqref{eq:rhol} replaced with $\pm$ to obtain the dynamics of the optical orientation $P_{\text{oo}}(t)$.

Figures~\ref{Fig5}(c,d) show the results of the calculation with the parameters $\alpha=1.5$, $\tau_0=4$~ns, $\Delta_B=10$~mT, $\tau_c=100$~ns, and $\varkappa=200$ after averaging over $10^7$ realizations of the random Hamiltonian~\eqref{eq:H} and with the additional depolarizing factor 0.5, which reflects the imperfect initialization of the exciton state~\cite{PhysRevB.96.035302}. One can see a remarkable agreement of the simulations and experimental results shown in Figures~\ref{Fig5}(a,b), which is achieved with reasonable parameter values, c.f. Tab.~\ref{tab:parameters}. Also the value of $\Delta_B=10$~mT corresponds to the energy scale of the hyperfine interactions in perovskite NCs ($1.2~\mu$eV)~\cite{hf1,hf2,hf3}.  For comparison, we show in Figures~\ref{Fig5}(e,d) the modeling of the spin dynamics without account of the hyperfine interactions with the nuclear spins. The difference from the modeling in Figures~\ref{Fig5}(c,d) is evident. For the optical alignment it occurs in zero and small magnetic fields and for the optical orientation in intermediate fields.  More details of the experimental dynamics and its modeling can be found in Fig.~\ref{SIcalc} in SI.

The polarization dynamics presented in Fig.~\ref{Fig5} can be understood as follows. At small time delays, the main contribution to the PL is provided by the excitons with short lifetimes and hence significant exchange splittings. The corresponding eigenstates are linearly polarized, see Fig.~\ref{Fig6}(a), so the optical alignment is significant in zero magnetic field, while the optical orientation is negligible. Accordingly, at short time delays, a large magnetic field of the order of one Tesla is needed to suppress the optical alignment and stabilize the optical orientation similarly to the direct excitons above the band edge.

By contrast, the PL at long time delays is governed by the SIE with long lifetimes and small exchange splittings. For them, the hyperfine interaction dominates the fine structure, Fig.~\ref{Fig6}(b). It leads to the formation of circularly polarized states with the exciton angular momentum projection $0,\pm1$ onto the direction of the random Overhauser field. The absence of optical alignment at the long time scales is the primary indication of the new regime, where the hyperfine interaction dominates in the exciton fine structure. In this regime, both the optical orientation and alignment are suppressed at zero magnetic field~\cite{random_triplet} (we remind that we also take into account the finite nuclear spin correlation time). However, a magnetic field as small as ten millitesla is sufficient to effectively decouple the SIE from the nuclear spins and to recover the optical orientation, see Fig.~\ref{Fig6}(c).

Notably, the proportionality constant $\varkappa$ between the exciton decay rate $1/\tau$ and the strength of the exchange interaction $\delta$ allows us to get insight into SIE localization potential. The exchange interaction can be estimated as
\begin{equation}
  \label{eq:Delta}
  \delta\sim\hbar\omega_{LT}a_B^3\int\Psi^2(\bm r,\bm r)\d\bm r,
\end{equation}
where $\hbar\omega_{LT}$ is the longitudinal-transverse splitting of exciton-polaritons in a bulk crystal, $a_B$ is the exciton Bohr radius, $\Psi(\bm r_e,\bm r_h)$ is the orbital part of the electron-hole envelope wave function in a NC (not to confuse with the spin part $\psi$ of the exciton wave function), and the integration runs over the coinciding electron and hole coordinates~\cite{BP}. The radiative decay rate can be estimated as
\begin{equation}
  \label{eq:gamma}
  \frac{1}{\tau}\sim\omega_{LT}(a_Bq)^3\left(\frac{3}{2+\eps_{NC}/\eps_b}\right)^2\left(\int\Psi(\bm r,\bm r)\d\bm r\right)^2,
\end{equation}
where $q=\omega_0\sqrt{\eps_b}/c$ is the wave vector of the light emitted at the frequency $\omega_0$ in the glass matrix with the dielectric constant $\eps_b$, while $\eps_{NC}$ is the dielectric constant of the perovskite. This expression takes into account the small size of NCs as compared to the light wave length.

Provided the electron and the hole are localized on the scale $l$ (smaller or comparable to the NC size) and the distance between them is $d$, the integrals in Eqs.~\eqref{eq:Delta} and~\eqref{eq:gamma} can be estimated as $\exp({-2d/l})/l^3$ and $\exp({-d/l})$, respectively. This gives the proportionality coefficient
\begin{equation}
  \varkappa\sim\frac{1}{(ql)^3}\left(\frac{2+\eps_{NC}/\eps_b}{3}\right)^2.
\end{equation}
Using the experimentally determined $\varkappa=200$, $\eps_b=3$~\cite{ANNAPURNA1991454}, and $\eps_{NC}=4.3$~\cite{pssr202300407}, we obtain the localization length of $l\sim13$~nm. This supports our interpretation of the SIE and shows that the distance between the localized electron and hole is indeed comparable to the size of the NCs, which we estimate as exceeding 16~nm.

Finally, we highlight that the dynamics of optical alignment and orientation cannot be described by the exciton exchange interaction alone, as we show in Fig.~\ref{Fig5}. Thus, the SIE represent the first platform to demonstrate the effect of the hyperfine interaction on the exciton fine structure and optical alignment effect in particular.

\section{Conclusions}
\label{sec:conclusions}

In conclusion, we have performed a comprehensive study of the magneto-optical properties of CsPbI$_3$ perovskite NCs, including measurements of the optical alignment and optical orientation of excitons under continuous-wave and pulsed excitation. Drastic differences of the exciton properties above and below the absorption edge are found. They are explained by formation of spatially indirect excitons in the large NCs. This leads to the power-law decay of the PL intensity, pronounced optical orientation in zero magnetic field, monotonous decay of the optical alignment and non-monotonous variation of the optical orientation with time. These observations are explained theoretically by considering Gaussian orthogonal ensemble of the exchange interaction Hamiltonians with the strength being inversely proportional to the exciton lifetime. For the strongly spatially indirect excitons, a fundamentally new regime, where the hyperfine interaction dominates the exciton fine structure, was experimentally realized for the first time.

For an outlook, we note that since the hyperfine interaction dominates in the fine structure of the most spatially indirect excitons, it can be used to control their magneto-optical properties. In particular, the fine structure can be manipulated by means of dynamic nuclear spin polarization and other nuclear spin tools. This is in contrast to the spatially direct excitons with the strictly fixed fine structure. The observed interplay between the exchange interaction and the Zeeman splitting in the polarization dynamics provides new insight into the exciton fine structure and spin physics of perovskite nanocrystals, which is important for their potential application in spin-optoelectronic and polarization-sensitive devices.

\section{Experimental sections}
\label{sec:Experimental}

\subsection{Sample}
The sample with CsPbI$_3$ NCs embedded in fluorophosphate glass Ba(PO$_3$)$_2$–AlF$_3$ was synthesized by rapid cooling of a glass melt enriched with the components required for the perovskite crystallization, using a technique similar to that described in Ref.~\cite{Kolobkova}.  The glass synthesis was performed in a closed glassy carbon crucible at the temperature of  $T = 1050^{\circ}$C.  The details of the sample preparation are  given in Ref.~\cite{meliakov2025}. The sample technology code is  EK34. The comparison of the PL spectrum of this sample with the spectra of samples EK31, EK7, and  EK8 with known dispersion of NC sizes~\cite{nestoklon2023} allows us to conclude that the characteristic diameter of the NCs in EK34 forming the absorption edge exceeds 16~nm.

\subsection{Optical absorption and photoluminescence}

The absorption spectrum was measured using a Cary 6000i spectrophotometer (Agilent Technologies). The sample was fixed on the cold finger of a flow-helium cryostat providing a sample temperature of 3.6~K. Since the perovskite samples exhibit highly efficient photoluminescence, two crossed polarizers and a diaphragm were installed in the spectrophotometer measurement channel to reduce the influence of the PL on the measured absorption  spectrum. Initially, the baseline spectrum was measured with all elements installed except the sample itself. The subsequent absorption spectrum measurements were performed using this baseline spectrum.

For PL measurements, the sample was mounted strain-free on a rotatable stage. It was immersed in pumped liquid helium at $T = 1.7$~K in a split-coil magnet cryostat and subjected to magnetic fields up to $B = 7$~T, which were applied parallel to the light wave vector (the Faraday geometry).

The PL emission was dispersed by a 0.5-m spectrometer. For the cw measurements the PL was detected by a liquid-nitrogen-cooled charge coupled device detector. The long-time PL dynamics at a selected spectral energy was detected by a GaAs photomultiplier operating in the time-correlated photon-counting mode. In order to monitor the PL decay over a wide temporal range up to 40~$\mu$s, the time resolution of the detection system was varied between 2.2~ns and 512~ns.  In these measurements, the photoluminescence was excited by a diode laser with a photon energy of 1.94~eV which operates in cw (with a power density of 0.01~W/cm$^2$) or pulse (with pulse width of 90~ps) mode.

\subsection{Optical alignment and orientation}

For measuring the optical alignment, the linear polarizations of the excitation laser and PL were selected by a Glan–Thompson prism and a half-wave plate. The linear polarization degree of the PL ($P_\text{l}$) induced under linearly polarized excitation is defined as
\begin{align}
  P_\text{l}= \frac{I^{\parallel}-I^{\perp}}{I^{\parallel} + I^{\perp}} ,
\end{align}
where $I^{\parallel/\perp}$ are the PL intensities, with the subscripts $\parallel/\perp$ corresponding to the parallel/perpendicular linear polarization directions of excitation and detection.

For the optical orientation measurement, the circular polarizations of the excitation laser and the PL were selected by  a Glan–Thompson prism and a quarter-wave plate. The optical orientation degree of the PL induced by circularly polarized excitation, $P_\text{oo}$, is defined as
\begin{align}
\label{eq:rho_c}
  P_\text{oo}= \frac{I^{co} - I^{cross}}{I^{co} + I^{cross}}.
\end{align}
Here, $I^{co/cross}$ is the intensity of the circularly co/cross-polarized PL components relative to the excitation.

\section*{Acknowledgements}
The Deutsche Forschungsgemeinschaft supported the work of N.E.K. (project KO 7298/1-1, no. 552699366) and D.R.Y. (project YA 65/28-1, no. 527080192). The experimental activities conducted by T.S.S., including investigation of the  magneto-optical  properties,  as well as the exciton recombination and the spin dynamics, were  supported  by a grant of the Russian Science Foundation (No. 22-12-00022-P). The theoretical modeling by D.S.S. was supported by the Russian Science Foundation Grant 25-72-10031.  M.S.K. and E.V.K. acknowledge support by the Saint Petersburg State University (Grant No. 125022803069-4).

\newpage


\setcounter{equation}{0}
\setcounter{figure}{0}
\setcounter{table}{0}
\setcounter{page}{1}
\setcounter{section}{0}
\makeatletter
\renewcommand{\thepage}{S\arabic{page}}
\renewcommand{\theequation}{S\arabic{equation}}
\renewcommand{\thefigure}{S\arabic{figure}}
\renewcommand{\thetable}{S\arabic{table}}
\renewcommand{\thesection}{S\arabic{section}}
\renewcommand{\bibnumfmt}[1]{[S#1]}
\renewcommand{\citenumfont}[1]{S#1}


\renewcommand{\S}{\mathop{\mathcal S}}
\newcommand{\X}{\mathop{\mathcal X}}
\newcommand{\Y}{\mathop{\mathcal Y}}
\newcommand{\Z}{\mathop{\mathcal Z}}




\begin{widetext}
\begin{center}
  \textbf{{\Large Supplementary Information:}}\\
  \textbf{{\Large Hyperfine versus exchange interaction in spin dynamics of spatially indirect excitons in CsPbI$_{3}$ perovskite nanocrystals}} \\~\\
{Timur~S.~Shamirzaev, Nataliia~E.~Kopteva, Dmitry~S.~Smirnov, Dmitri~R.~Yakovlev, Elena~V.~Kolobkova, Maria~S.~Kuznetsova, Eugeniyus~L.~Ivchenko, Yan E. Maidebura, Evgeny~A.~Zhukov,  and Manfred~Bayer}\\~\\~
\end{center}
\end{widetext}

\section{Additional experimental data}

For the highest energy PL contribution at 1.739~eV, the shape of the band reflects the distribution of the NCs with sizes smaller than twice the Bohr radius. The middle band with the energy maximum of 1.708 eV corresponds to direct exciton recombination in NCs with  sizes of the order or larger than twice the exciton Bohr radius of about 10~nm.  In these NCs the excitons are quantized in their center of mass coordinate, however, the quantization energy is small. This explains the relatively small width of this band.

The short-time PL dynamics was measured with a Hamamatsu streak camera coupled to a 0.5 m spectrometer.  The spectral resolution was typically about 1~meV and the time resolution was about 30~ps. In these measurements, the sample was excited by a pulsed (0.1~ps) laser with a photon energy of 2.53~eV.

\begin{figure}[hbt]
\centering
\includegraphics[height=5.5cm]{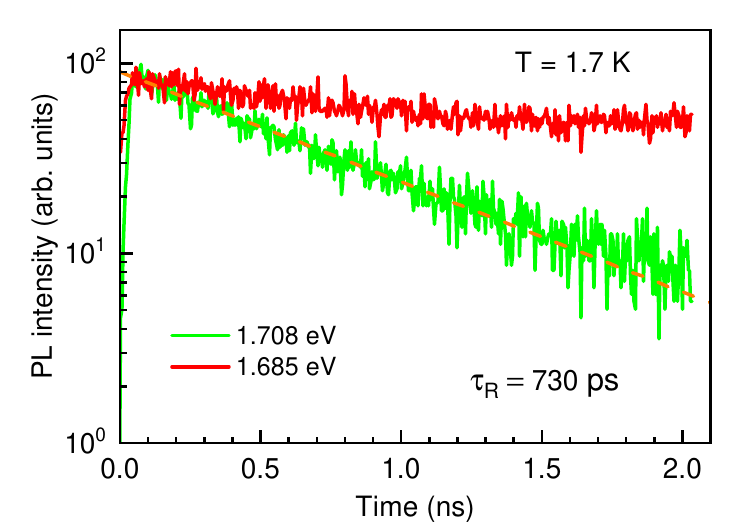}
\caption{\label{SIF0} Fast PL dynamics  measured at detection the energies of 1.685~eV (red) and 1.708~eV (green). The orange dashed line shows an exponential fit with lifetime $\tau_R =730$~ps.}
\end{figure}

One can see in Fig.~\ref{SIF0} that above the absorption edge at the energy of 1.708~eV, the dynamics demonstrates a mono-exponential decay in the subnanosecond time range, which is typical for exciton recombination in lead halide perovskite NCs. However, when the detection energy decreases below the absorption edge (1.685~eV), the dynamics becomes non-exponential following the $I(t) \propto t^{-\alpha}$ dependence.

\section{Modeling details}

\begin{figure}[hbt]
\centering
\includegraphics* [height=5.5cm]{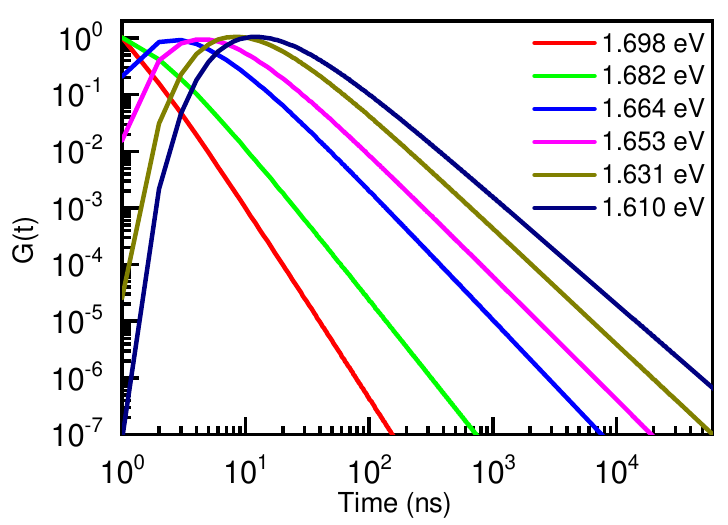}
\caption{\label{SIF1} Distributions of exciton lifetimes, $G(\tau)$, with the parameters presented in Table~\ref{tab:parameters}, corresponding to NC subensembles emitting at different energies.}
\end{figure}

Figure~\ref{SIF1} shows the distribution functions $G(\tau)$ for different NC subensembles, obtained by fitting of the recombination dynamics shown in Fig.~\ref{Fig1}(c) using Eq.~\eqref{eq1}.

\clearpage

\section{Details of polarization dynamics}

\begin{figure}[hbt]
\centering
\includegraphics*[width=\linewidth]{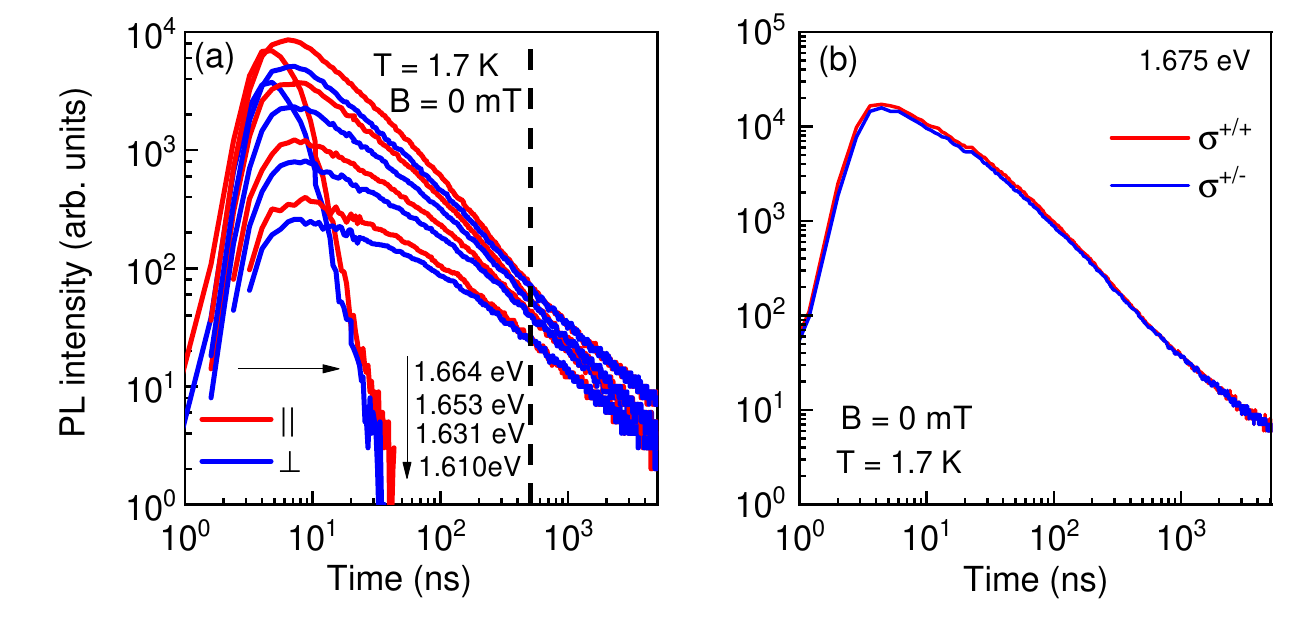}
\caption{\label{SIF2} (a) PL dynamics measured in the polarization parallel ($\parallel$, red) and perpendicular ($\perp$, blue) to the linear polarization of excitation. (b)~PL dynamics measured in co- ($\sigma^{+/+}$, red) and  cross- ($\sigma^{+/-}$, blue) circular polarizations for $\sigma^+$-polarized excitation.}
\end{figure}

The PL dynamics measured in linear polarization $\parallel$ and $\perp$  to the linearly polarized excitation at different detection energies in zero magnetic field is shown in Fig.~\ref{SIF2}(a). In the spectral range below the absorption edge, the linear polarization of the emission disappears at about 500~ns after the excitation pulse (shown by the black dashed line), regardless of the total PL decay duration.

The PL dynamics measured in $\sigma^+$ and $\sigma^-$ circular polarization for $\sigma^+$ polarization of excitation at zero magnetic field is shown in Fig.~\ref{SIF2}(b).

\begin{figure}[hbt]
\centering
\includegraphics*[height=6cm]{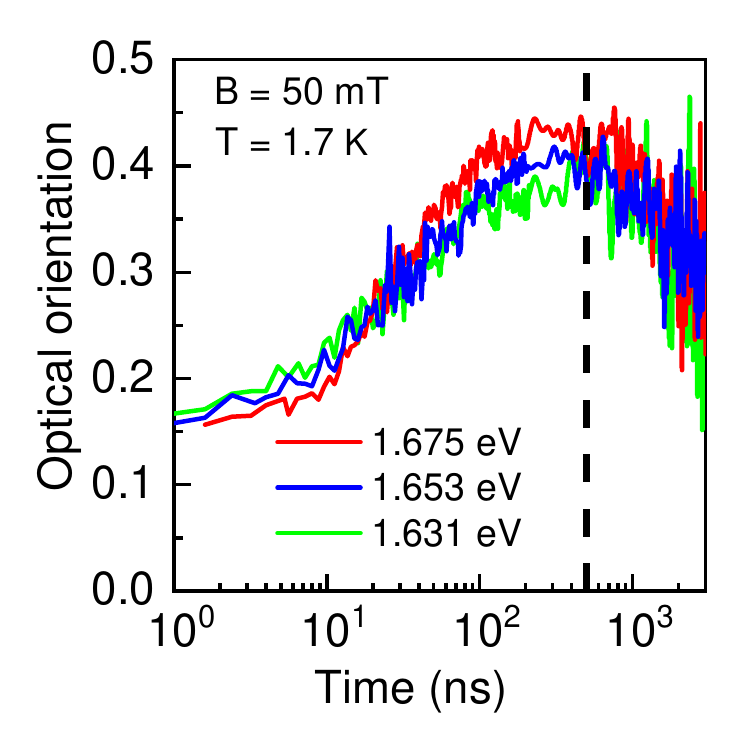}
\caption{\label{SIF3}  Dynamics of the optical orientation degree measured in the magnetic field of 50~mT at the detection energies of 1.675~eV (red), 1.653~eV (blue), and 1.631~eV (green). The dashed line marks the delay of 500~ns, where the circular polarization degree reaches its maximum. }
\end{figure}

It is interesting that the dynamics of optical orientation in a longitudinal magnetic field of 50~mT shows a nonmonotonous behavior in the spectral range ($1.631 - 1.675$~eV), below the absorption edge, see Fig.~\ref{SIF3}. Immediately after the end of the excitation pulse, the polarization degree is about 16$\%$. Then it increases, reaching a maximum of about 40\% at delays of about 500~ns, see the dashed line in Fig.~\ref{SIF3}.

\clearpage
\section{Optical orientation and optical alignment for cw emission}

\begin{figure}[hbt]
\centering
\includegraphics*[height=13cm]{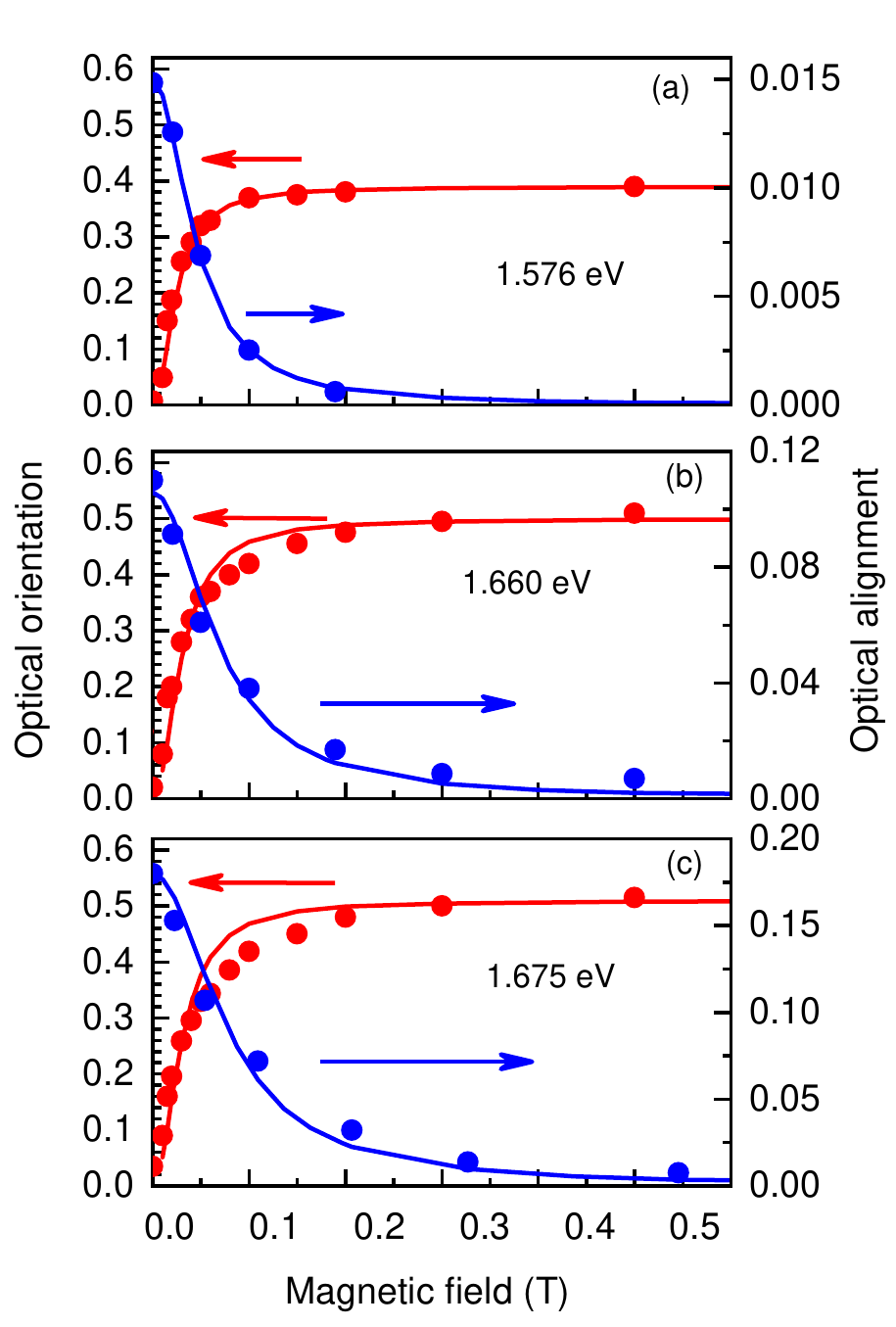}
\caption{\label{SIF4} Optical orientation and optical alignment for cw excitation vs longitudinal magnetic field measured at energies: (a) 1.576 eV, (b) 1.660 eV, and (c) 1.675 eV.
}
\end{figure}

The magnetic field induced changes in optical alignment and orientation at the detection energies $1.576$~eV, $1.660$~eV, and $1.675$~eV are shown in Figs.~\ref{SIF4} (a),~\ref{SIF4}(b), and~\ref{SIF4}(c), respectively.
The values of $\delta_{\rm l}$, $\delta_{\rm oo}$, $P^{\text{sat}}_\text{oo}$, and $P_{\text{l}}^0$ that are obtained from the best fit of the dependences $\rho_\text{l}(B)$ and $\rho_\text{oo}(B)$  by Eq.~\eqref{eq:a1} and Eq.~\eqref{eq:a2}  are collected in Table~\ref{tab:HWHM}.

\clearpage

\section{Theoretical details}

\begin{figure}[hbt]
\centering
\includegraphics[width=0.8\linewidth]{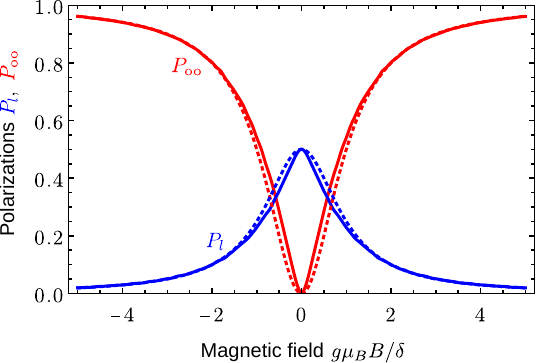}
\caption{\label{fig:cf} Dependences of the optical alignment (blue lines) and the optical orientation (red lines) on the longitudinal magnetic field, calculated after Eqs.~\eqref{eq:long} (solid lines) and~\eqref{eq:a} (dotted lines).
}
\end{figure}

The dependences of the optical orientation and alignment on the longitudinal magnetic field in the limit of long exciton lifetimes accounting for the exchange interaction and neglecting the hyperfine interaction can be calculated after Eqs.~\eqref{eq:long}. In Fig.~\ref{fig:cf} we show that the result agrees well with the numerical approximations~\eqref{eq:a} (we set here $P^{0}_\text{l}=0.5$ and $P^{\text{sat}}_\text{oo}=1$).

We stress that for the given NCs there are two random splittings between the exciton states, as explained above. Therefore, in each NC, the intensities of the emitted light in two arbitrary polarizations are determined by two Lorentzians with the corresponding widths and the amplitudes determined by the Euler angles, as mentioned above. However, after averaging over all these parameters in an ensemble of NCs, the expressions become similar to Eq.~\eqref{eq:a}. We stress that these equations are not exact, they are numerical approximations only. The parameter $\delta$ here is exactly the same parameter that determines the dispersion of the components of the tensor $\hat{\bm \Delta}$. This tensor of the exchange splittings $\hat{\bm\Delta}$ can also include contributions from low symmetry of a perovskite crystal phase. Since in the experiment the widths of the magnetic field dependences of the optical orientation and alignment are different, we denote their values determined from the fits as $\delta_{\rm oo}$ and $\delta_{\rm l}$, respectively.

We remind that $\tau$ denotes the exciton lifetime both in the experimental and theoretical parts of the paper. We denote by capital $X$, $Y$, and $Z$ the exciton states, while the subscripts $x$, $y$, and $z$ denote the components of the exciton wave functions. In Eq.~\eqref{eq:Ixy_tau}, the squared components of the wave function determine the occupancies of the corresponding states, while the probability of photon emission is proportional to $1/\tau$. Since $\tau$ is different in different NCs, it is important to keep it in this equation.

\begin{figure*}[hbt]
\centering
\includegraphics[width=\linewidth]{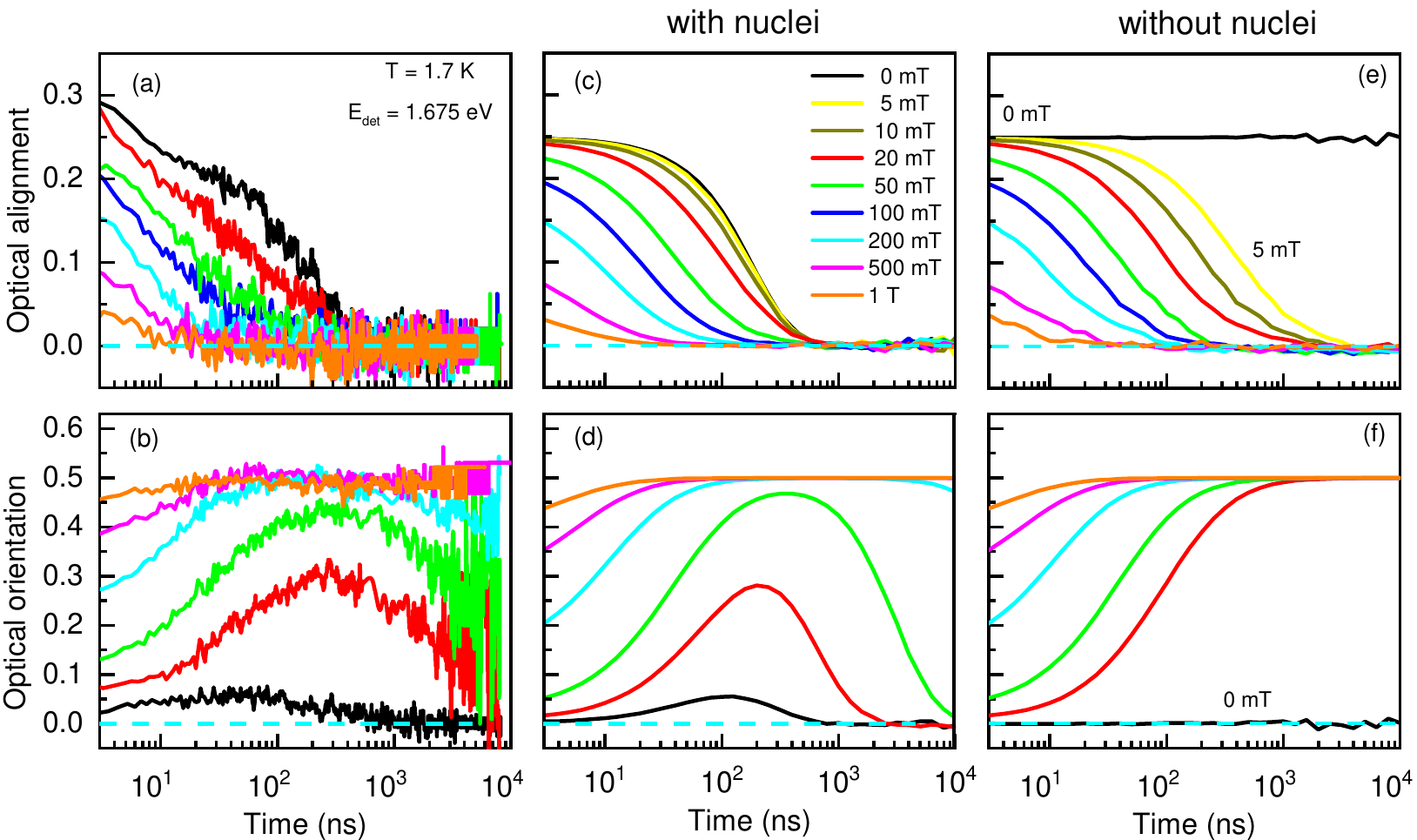}
\caption{\label{SIcalc}  Dynamics of optical alignment (a)  and  optical orientation (b) measured at the energy of 1.675~eV in different magnetic fields. Model calcualtions of the polarization dynamics are shown with account of the hyperfine interaction of the SIE with nuclear spins (c,d) and without it (e,f). The model parameters are given in the main text. These dependences are complementing the data presented in Fig.~\ref{Fig5} of the main text.
}
\end{figure*}

Figure~\ref{SIcalc} shows the same as Fig.~\ref{Fig5} but for more different magnetic fields. The main evidence for the role of the hyperfine interaction is the absence of the optical alignment of excitons at long times. It points to spin relaxation of the localized excitons. The primary candidate for this is the hyperfine interaction, which has also the right corresponding energy scale. The absence of optical orientation also points out to spin relaxation, which is already known to be driven by the hyperfine interaction in perovskite nanostructures in small magnetic fields. This regime was theoretically predicted for doublet excitons in (In,Al)As quantum dots, and was never reached before experimentally to the best of our knowledge.

The parameters of the model calculations are kept constant for all magnetic fields. Note that in the absence of the exchange and Zeeman interactions, the state with zero angular momentum projection also depends on the direction of $\bm\Omega_N$. The strength of the hyperfine interaction does not depend on the temperature as long as the electron and hole remain spatially indirect, and the nuclear spin fluctuations are zero on average. However, the temperature can affect the crystal phase of NCs, and thus change the exchange splittings. Nevertheless, the exchange interaction requires overlap between the electron and hole wave functions, so it remains small. On the other hand, with increasing temperature, the phonon assisted transitions between exciton levels may come into play. As soon as the corresponding rate becomes faster than the nuclei induced exciton spin precession, the hyperfine structure of the exciton levels cannot be resolved anymore.

Note that Eq.~\eqref{eq:Delta} is valid for both SIE and direct excitons. The estimation of $l$ does not involve directly parameters such as $\delta$ ($\delta_{\rm l}$ or $\delta_{\rm oo}$), it is based mainly on the value of $\varkappa$, which is determined from fitting the dynamics of optical orientation and alignment shown in Fig.~\ref{Fig5}.

\end{document}